\documentclass[aps,prb,reprint,amsmath,amssymb,showpacs]{revtex4-1}
\usepackage{xcolor,color}
\usepackage{bm,graphicx,braket}
\newcommand*\ii{\mathrm{i}}
\newcommand*\ee{\mathrm{e}}
\usepackage{hyperref}

\begin{document}

\title{Real-space cluster dynamical mean-field theory:\\ Center focused extrapolation on the one- and two particle level}

\author{Marcel Klett$^1$}
\email{marcel.klett@uni-tuebingen.de}
\author{Nils Wentzell$^2$}
\author{Thomas Sch\"afer$^{3,4}$} 
\author{\mbox{Fedor Simkovic IV$^{3,4}$}}
\author{Olivier Parcollet$^{2,5}$}
\author{Sabine Andergassen$^{1}$}
\author{Philipp Hansmann$^{6,7}$}
\affiliation{$^1$Institut f\"ur Theoretische Physik and Center for Quantum Science, Universit\"at T\"ubingen, Auf der Morgenstelle 14, 72076 T\"ubingen, Germany}
\affiliation{$^2$Center for Computational Quantum Physics, Flatiron institute,
Simons Foundation, 162 5th Ave., New York, 10010 NY, USA}
\affiliation{$^3$Coll{\`e}ge de France, 11 place Marcelin Berthelot, 75005 Paris, France}
\affiliation{$^4$CPHT, CNRS, {\'E}cole polytechnique, IP Paris, F-91128 Palaiseau, France}
\affiliation{$^5$Universit\'e Paris-Saclay, CNRS, CEA, Institut de physique th\'eorique, 91191, Gif-sur-Yvette, France}
\affiliation{$^6$Max-Planck-Institut f\"ur Chemische Physik fester Stoffe, N\"othnitzerstr. 40, 01187 Dresden, Germany}
\affiliation{$^7$Department of Physics, University of Erlangen-Nuremberg, 91058 Erlangen, Germany}

\begin{abstract}
  We revisit the cellular dynamical mean-field theory (CDMFT) for the single
  band Hubbard model on the square lattice at half filling, reaching real-space
  cluster sizes of up to $9\times9$ sites.  
  Using benchmarks against direct lattice diagrammatic Monte Carlo
  at high temperature, we show that the self-energy obtained
  from a cluster center focused extrapolation converges faster with
  the cluster size than the periodization schemes previously
  introduced in the literature.  The same benchmark also shows that
  the cluster spin susceptibility can be extrapolated to the exact
  result at large cluster size, even though its spatial extension is
  larger than the cluster size.
\end{abstract}
\maketitle

\section{Introduction}
Even after decades of intense research, the single band Hubbard model
in finite dimensions larger than one remains an unsolved cornerstone
paradigm in theoretical solid state physics. As a non-perturbative
technique dynamical mean-field theory (DMFT)~\cite{DMFT,DMFT2, DMFTreview} led to
a significant leap forward as it provided an approximation which is
nowadays widely used to treat material realistic lattice models~\cite{KotliarRMP2006}. 
In this approximation, the lattice problem is mapped to 
a self-consistent auxiliary quantum impurity model, leading to a local approximation of the 
self-energy.

Single site DMFT has led to many successes, e.g. a description of the
Mott-Hubbard metal-to-insulator transition (MIT) from a weak coupling metallic
over to a strongly correlated Fermi liquid and eventually Mott insulating
state~\cite{DMFTreview,V2O3review} within a single theoretical framework. The
locality of the self-energy is however a major limitation for some of the most interesting
problems as for example cuprate high-$T_c$ superconductivity~\cite{bednorz,dagotto,timusk,lee}
where the pseudo-gap phase is characterized by a strong node-antinode
differentiation. 
Note, however, that in material-realistic multi-orbital models even local DMFT self-energies
can lead to $\mathbf{k}$ dependent effects in the spectral function $A(\mathbf{k},\omega)$
due to orbital mixing as, e.g., shown in \cite{PhysRevX.9.021048}.
Quantum critical
behavior~\cite{loehneysen,buettgen,gegenwart} in low dimensions and/or at low
temperatures is another example where the physics is dominated by non-local
correlations beyond the mean-field description of spatial fluctuations within
DMFT.

In recent years, several methods have been introduced to overcome the
locality of the self-energy approximation in DMFT.  Firstly, cluster
extensions of DMFT, in reciprocal space (DCA) \cite{DCAOriginal, DCA2,
  DCA1, cluster_rev}, in real space (CDMFT) \cite{CDMFTOriginal1,
  CDMFTOriginal} or in its first "nested" form \cite{DMFTreview,
  causality1, JacksaLuttWard}.  This transformed DMFT into a
controlled method.  The control parameter is the size of the cluster,
which determines also the resolution of the approximation in momentum
space.  These approaches have found extensive use, e.g. in the study
of the Hubbard model
\cite{olivier-CDMFT,mottness,sakai-CDMFT,park-CDMFT,
   2x2-tremblay2,PhysRevLett.122.067203,2x2-tremblay3,2x2-tremblay,2x2-Nils, HubbardCluster1, HubbardCluster2,HubbardCluster3,HubbardCluster4,HubbardCluster5,HubbardCluster6,HubbardCluster7,HubbardCluster8, Ferrero_2009}.
Secondly, several diagrammatic
extensions~\cite{DGA1,DGA1b,DGA2,DGA3,DF1,DF2, TRILEX1, TRILEX2,
  TRILEX3, TRILEX4, TRILEX5,
  1PI,FLEX,Multiscale2,Multiscale,DMF2RG,Rohringer-Review} have been
proposed to obtain a better resolution in momentum or a better
convergence than cluster methods.

\begin{figure}[t]
  \centering
  \includegraphics[width=0.4\textwidth]{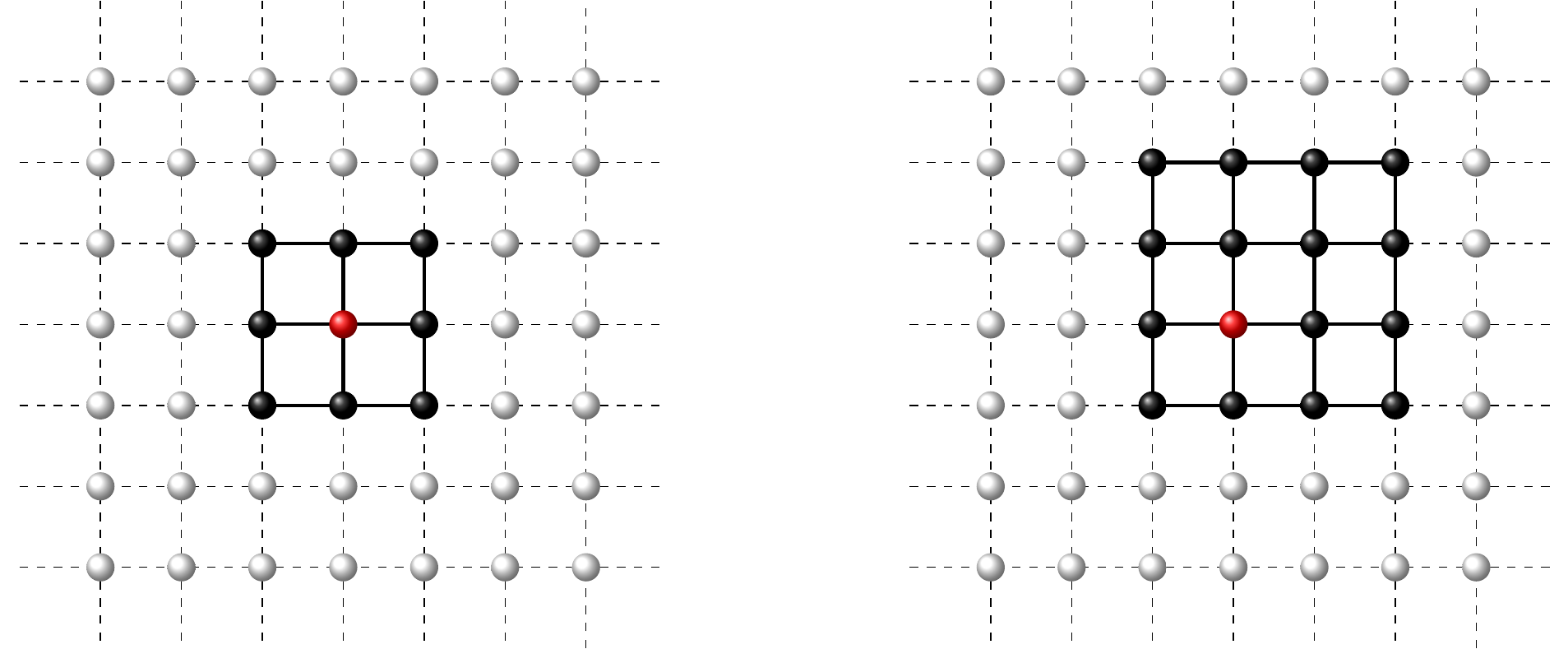}
  \caption{Square clusters with $N\times N$ lattice sites. The left
    panel shows the $3 \times 3$ and the right one the $4\times 4$
    cluster, with the central site ${\bf R}_c$ in red.}
  \label{fig:sketch}
\end{figure}

In this paper, we focus on the real space CDMFT method  \cite{CDMFTOriginal}.
While large cluster sizes were studied in DCA up to
convergence and compared to exact methods like diagrammatic Monte Carlo
in some parameter regimes, e.g. \cite{PhysRevB.96.041105, LeblancPRX}, large CDMFT clusters
have not received the same attention.  The central issue is that CDMFT breaks
translational invariance by definition for any finite cluster.  One therefore
needs to use re-periodization schemes to restore the translation
symmetry of the lattice self-energy when approximating it from the cluster
self-energy \cite{CDMFTOriginal}.  Several ways to do this have been discussed
in the literature \cite{PhysRevB.74.125110,PhysRevLett.84.522, sakai-CDMFT}.

In this work, we solve large CDMFT clusters of up to $9\times 9$ sites
for the half-filled Hubbard model, and benchmark their convergence
against exact results obtained with diagrammatic Monte Carlo (DiagMC)
\cite{Prokofev1998,VanHoucke2010,kozik2010diagrammatic} in its connected
determinant formulation (CDet~\cite{CDET2} for one particle reducible
quantities and $\Sigma$DDMC \cite{CDET4,CDET3,CDET} for one particle irreducible quantities).

We show that if we use a {\it center-focused extrapolation} (CFE) to
approximate the lattice self-energy, instead of averaging over the
whole cluster as done in previous works, we obtain a much better
approximation. This CFE converges faster and is in excellent agreement
with the DiagMC result at moderate temperature.  We further present
results at lower temperature $T$ and larger $U$ than the range
currently accessible with diagrammatic Monte Carlo methods.

The paper is organized as follows. In Section~\ref{sec:model} we
introduce the 2D Hubbard model and present the CDMFT formalism.  We
then present in Section~\ref{sec:CFE} the center focused
extrapolation technique, and benchmark it against DiagMC. In
Section~\ref{sec:spinsusc} we discuss the spin-spin correlation
function and finally summarize our results in
Section~\ref{sec:conclusion}.

\section{Model and Method}
\label{sec:model}

We study the 2D single-orbital Hubbard model on the square lattice
\begin{equation}
  \label{eq:hamiltonian}
  H = - t \sum_{<i,j>,\sigma} c_{i,\sigma}^{\dagger} c_{j,\sigma}  + \sum_{i} U \hat{n}_{i,\uparrow} \hat{n}_{i,\downarrow} - \mu \sum_{i,\sigma} \hat{n}_{i,\sigma}\;,
\end{equation}
where $c_{i,\sigma}^\dagger$ ($c_{i,\sigma}$) denotes the creation
(annihilation) operator for an electron with spin $\sigma$ on site
$i$, with the density operator
$\hat{n}_{i,\sigma} = c_{i,\sigma}^{\dagger} c_{i,\sigma}$ and the
chemical potential $\mu$. $t$ and $U$ are nearest-neighbor hopping
amplitude and onsite Hubbard repulsion respectively. We study the
half-filled case which corresponds to $\mu = U/2$.  For a
single-orbital unit cell the electronic dispersion relation in
momentum space reads
$\varepsilon(\mathbf{k}) = -2 t ( \cos k_x + \cos k_y)$.

In the following, we compute propagators of single- and two-particle
excitations in the CDMFT approximation \cite{CDMFTOriginal}.
CDMFT is a DMFT approximation on the super-lattice made of 
real space clusters on the original Bravais lattice \cite{CDMFTOriginal}, 
where the cluster sites within the super-cell play the role of orbitals.
Hence, the Green's functions and self-energy contains inter-sites elements.
They are computed using a self-consistent quantum impurity model, 
which is solved by a continuous time Quantum Monte-Carlo solver using an interaction expansion
\cite{Rubtsov2004}, implemented with the TRIQS library ~\cite{triqs_lib}.

For the present study we consider square clusters of up to $N \times N=9\times 9$ atoms
(see Fig.~\ref{fig:sketch} for a schematic illustration). Technically, the
equations for the Green's functions and self-energies are identical to those of
a multi-orbital (single-site) DMFT with a density-density interaction. The local
lattice Green's function in Matsubara frequencies is computed by integrating
over the reduced Brillouin zone (RBZ) of the super-lattice :
\begin{equation}
  \label{eq:gloc}
  G^\text{loc}_{ij}(\ii \omega_n) = \sum_{\mathbf{k}\in RBZ} \left[(\ii \omega_n + \mu)\delta_{ij} - \tilde{\varepsilon}_{ij}(\mathbf{k}) - \Sigma_{ij}(\ii \omega_n)\right]^{-1}\;,
\end{equation}
with $i,j$ indexing cluster sites, dispersion relation
$\tilde{\varepsilon}_{ij}(\mathbf{k})$ (with
$\mathbf{k} \in {\rm RBZ}$), and self-energy
$\Sigma_{ij}(\ii \omega_n)$. 
The Weiss field is given by the standard equation \cite{CDMFTOriginal}
\begin{equation}
  \label{eq:weiss-field}
  \left[\mathcal{G}^{0}(\ii \omega_n)\right]^{-1}_{ij} = \left[G^\text{imp}(\ii \omega_n)\right]^{-1}_{ij} + \Sigma_{ij}^{imp}(\ii \omega_n)\;.
\end{equation}
The self-energy acquires intersite components. Hence it can capture correlation effects
up to distances of about the linear size of the cluster.

\begin{figure*}
  \centering
  \includegraphics[width=0.75 \textwidth]{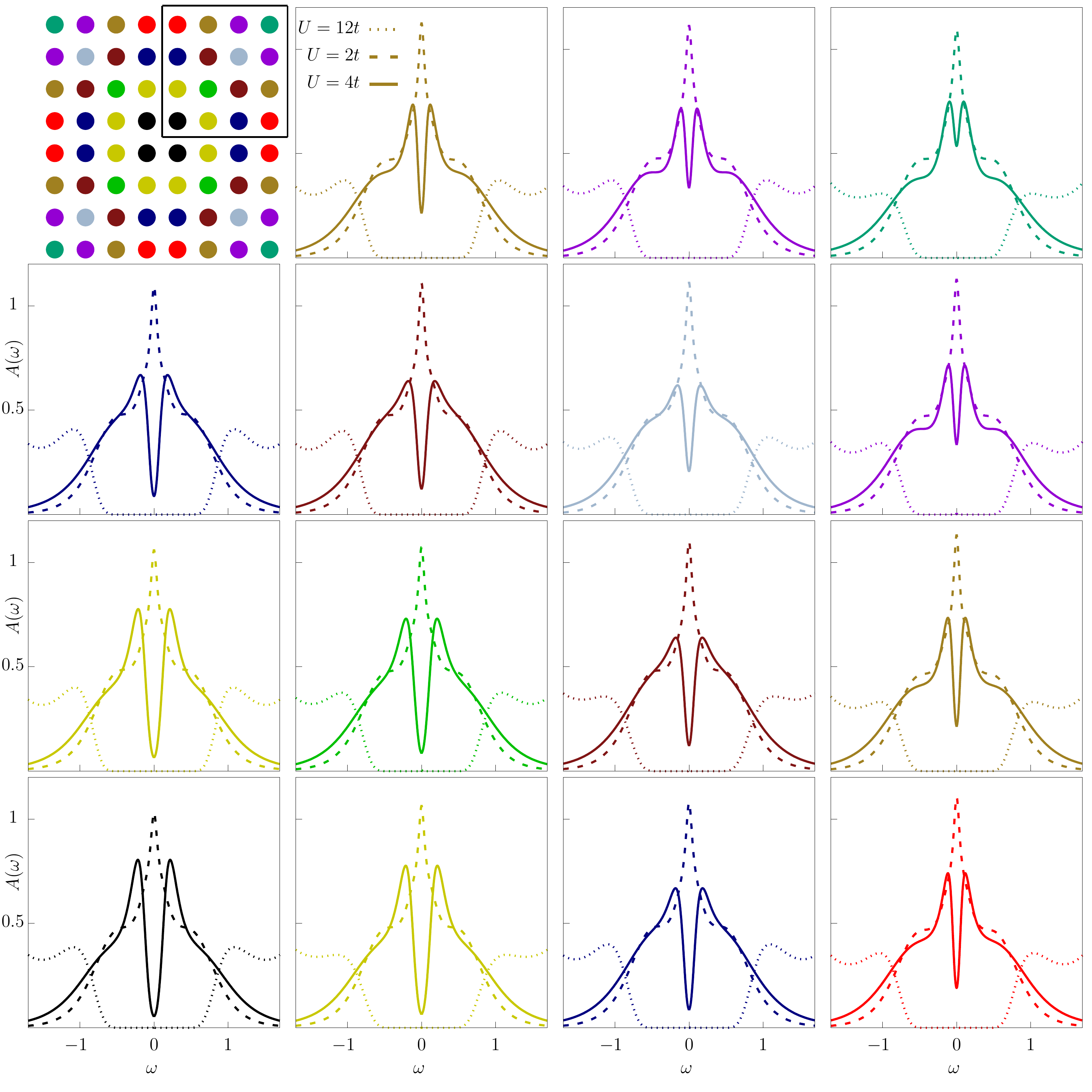}
  \caption{Local spectral functions for $\beta t = 12.5$,
    $N \times N=8 \times 8$ and different values of the interaction strength
    $U$, see upper left panel for the color coding of the cluster
    sites. There is a clear trend towards insulating behavior
    as we traverse the cluster from the border to the center.}
  \label{fig:cluster-A}
\end{figure*} 

\begin{figure}[h]
  \centering
  \includegraphics[width=0.48 \textwidth]{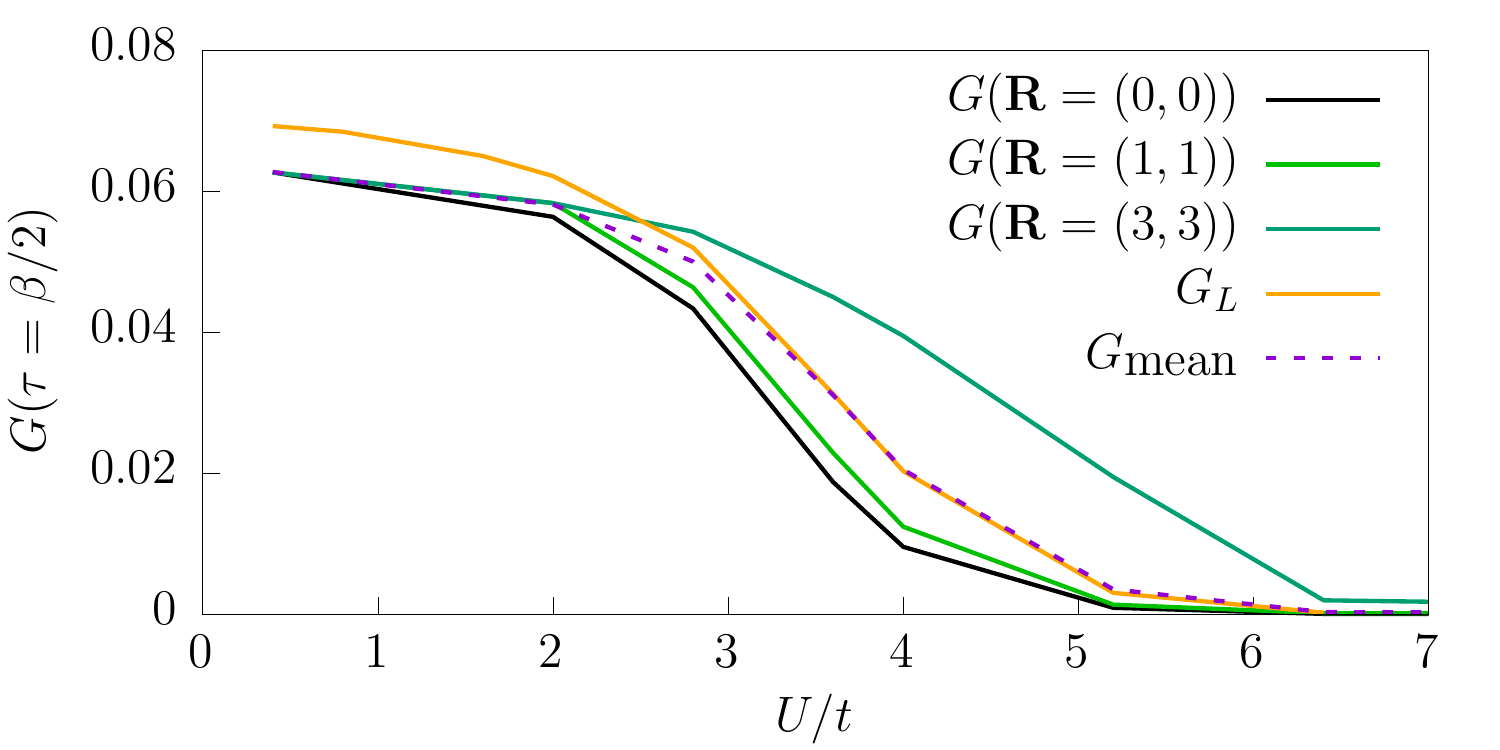}
  \caption{Averaged (dashed) and site dependent $G(\tau = \beta/2)$
    as a function of $U/t$ with parameters identical to Fig.~\ref{fig:cluster-A}.
    $G_{loc}$ is the local lattice Green's function obtained by using the cumulant periodization.}
  \label{fig:mit-lattice-site}
\end{figure}

The cluster self-energies break translational invariance for any finite cluster size.
Transformation to lattice quantities in the Brillouin zone of the original lattice therefore requires
restoring translational invariance for the Green's function, the self-energy, or its
cumulants~\cite{CDMFTOriginal1, CDMFTOriginal, mottness,PhysRevB.74.125110,PhysRevLett.102.056404,PhysRevB.88.115101,sakai-CDMFT,PhysRevLett.116.057003}
(see also Appendix~\ref{app:reper}).

\begin{figure*}
  \centering
  \includegraphics[width=0.902\textwidth]{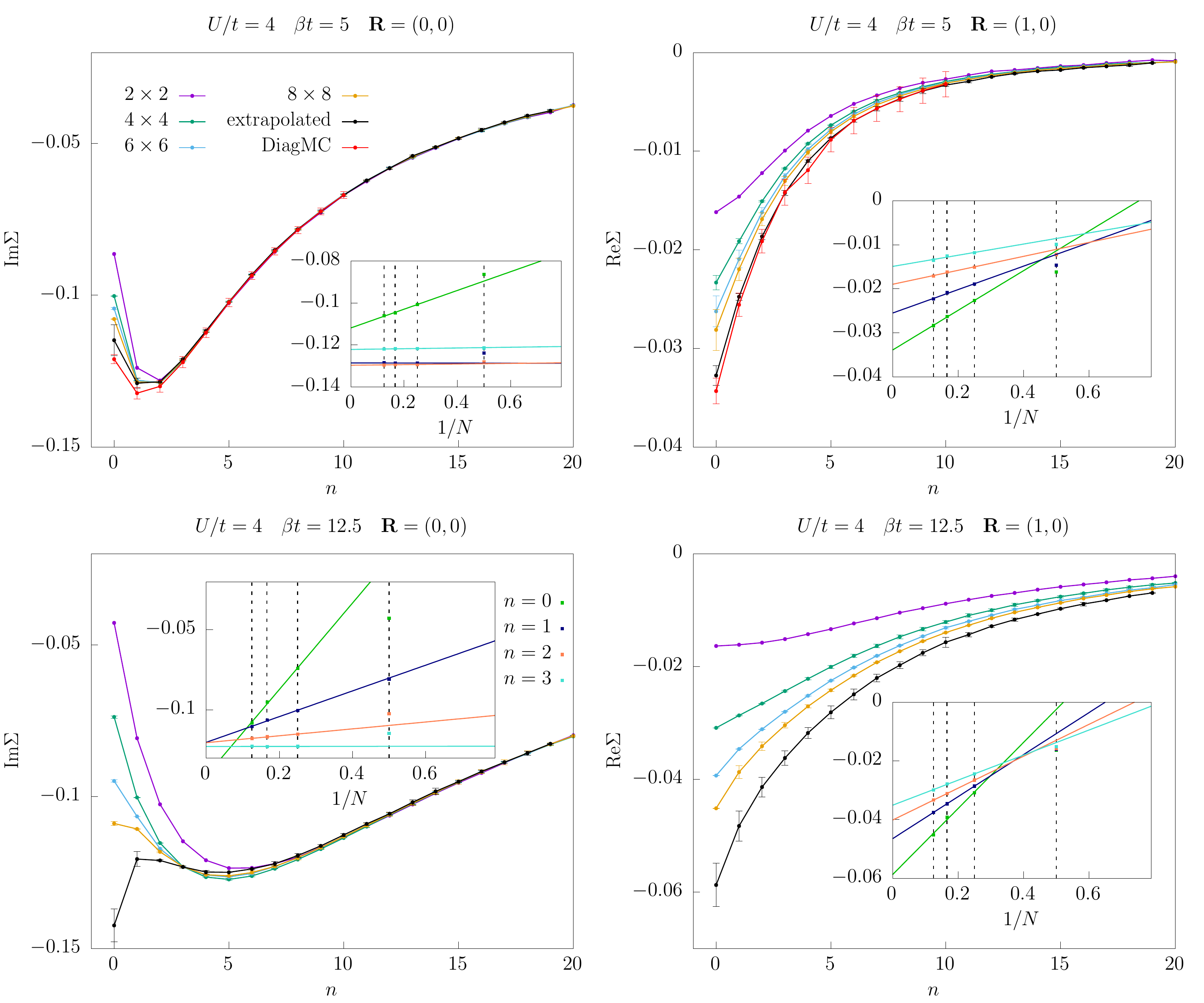}
  \caption{Matsubara frequency dependent self-energy of the central
    site (left) at $\mathbf{R}=(0,0)$ and of the neighboring one (right) at
    $\mathbf{R}=(1,0)$, as obtained from CDMFT and DiagMC calculations for
    $U/t=4$ and different temperatures. The extrapolated self-energy
    is obtained by fitting the data with a function linear in $1/N$
    (excluding the $2 \times 2$ data), see insets for the fits from
    the first four Matsubara frequencies.}
  \label{fig:SigmaR00}
\end{figure*}

\begin{figure}
  \centering
  \includegraphics[width=0.49\textwidth]{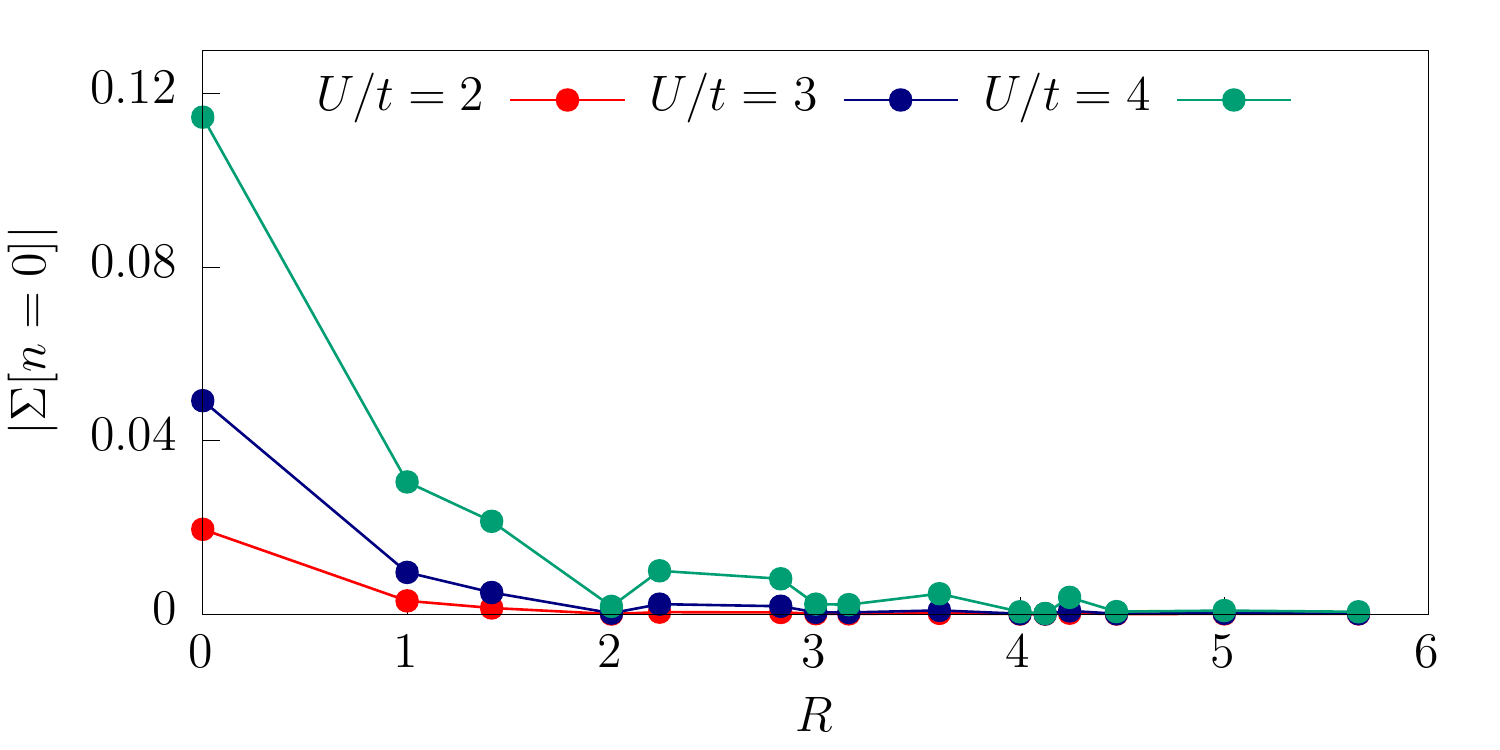}
  \caption{Absolute value of the self-energy at the first Matsubara
    frequency $\Sigma(i\omega_1,\mathbf{R})$ as a function of real-space
    distance $R=|\mathbf{R}|$ at $\beta t = 5$ and different values of the interaction
    $U/t$.}
  \label{fig:self_energy_vs_R}
\end{figure}
\begin{figure*}[ht]
  \centering
  \includegraphics[width=0.99\textwidth]{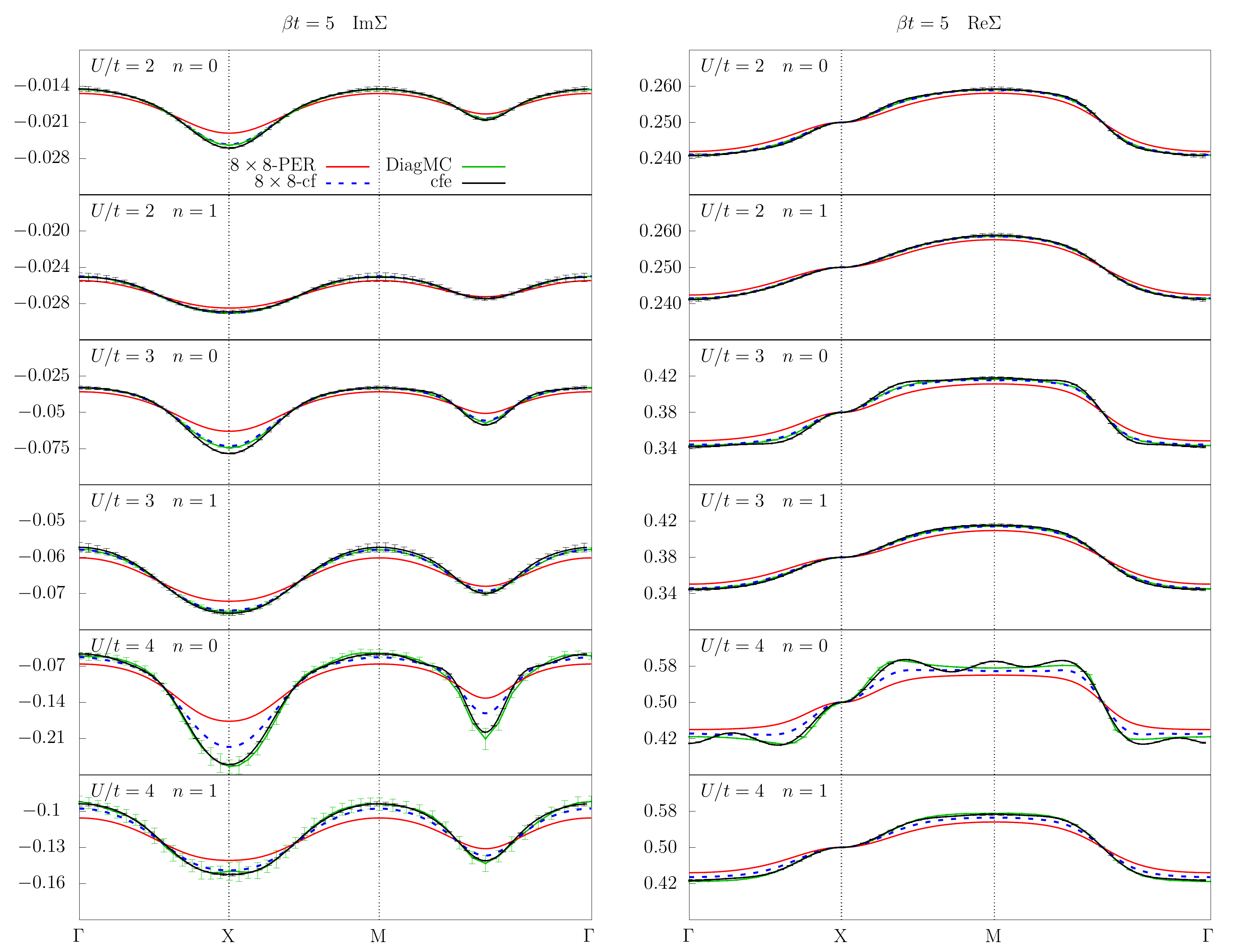}
  \caption{Imaginary (left) and real (right) part of the lattice self-energy
    $\Sigma(\mathbf{k})$ at the first and second Matsubara frequency $n=0,1$
    along a path in the first Brillouin zone, for
    an inverse temperature $\beta t =5$ and different values of the
    interaction $U/t$. We plot the lattice self-energy obtained by
    conventional periodization of the $8 \times 8$ cluster (red),
    Fourier transformation of the real-space correlations relative to the
    central site for the $8 \times 8$ cluster (blue), Fourier transformation
    using the extrapolated real-space self-energy (black),
    and the numerically exact DiagMC data (green).}
  \label{fig:CDET-CDMFT}
\end{figure*}

\begin{figure}
  \centering
  \includegraphics[width=0.45\textwidth]{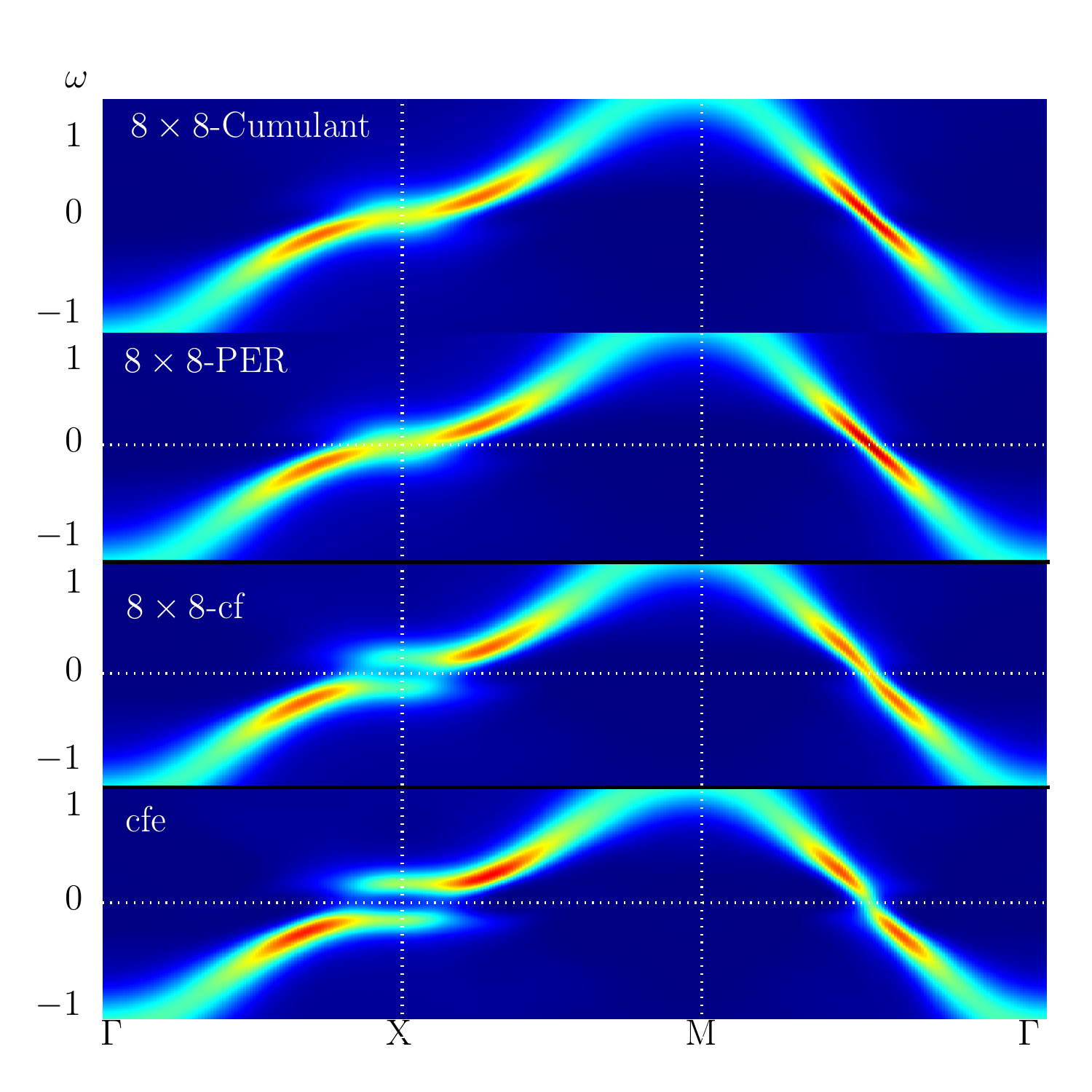}
  \caption{Spectral function of the lattice Green's function, for
    $U/t=4.0$ and an inverse temperature $\beta t = 5$, using the
    $8 \times 8$ cumulant periodization scheme, the conventional
    $8 \times 8$ periodization, the $8 \times 8$ center focused
    re-periodization and the center focused extrapolation (from top
    to bottom).}
  \label{fig:compare-spectra}
\end{figure}

\section{Center focused extrapolation}
\label{sec:CFE}

In a very large cluster, we expect on general grounds the Green function at the
center of the cluster to converge faster with cluster size than at the boundary. In this
section, we use this insight to improve on the re-periodization
within the CDMFT algorithm. We begin with the study of single-particle cluster quantities, and
analyze the convergence of the local and the nearest neighbor self-energy w.r.t.~cluster
size. In a second step we use these converged self-energies to approximate the lattice
self-energy, and benchmark it against exact DiagMC results.

\subsection{Cluster quantities}
\label{sec:realspace}

Let us first concentrate on the cluster Green's and spectral functions.
In Fig.~\ref{fig:cluster-A} we show site-resolved single particle
spectral functions obtained from an $N\times N=8 \times 8$ CDMFT calculation
for different values of $U$. Exploiting cluster symmetry, we show only the
upper right quadrant of the $64$ cluster sites, where equivalent
spectra are plotted in the same color. The violation of the
translational invariance is most pronounced for $U/t=4$ (solid lines)
as can be seen by comparing the spectra in the center at $\mathbf{R}=(0,0)$
with those in the corners at $\mathbf{R}=(3,3)$. For $U/t=2$ all spectra exhibit
a finite quasiparticle weight at $\epsilon_F$, while for $U/t=12$ they
are all gaped.

In Fig.~\ref{fig:mit-lattice-site} we show 
the corresponding site-dependent value of $G(\mathbf{R}, \tau=\beta/2)$.
The spectra in the cluster center are generally more correlated/insulating than
those at the edges/corners of the cluster. 
Given the fact that the exact solution has a critical value of $U_c=0$
at $T=0$, the single particle spectrum of the central site appears to
be a better approximation than the average over the whole cluster (dashed line in Fig.~\ref{fig:mit-lattice-site}).

\subsection{Convergence of cluster quantities}
\label{sec:conv_cluster_qties}

We now study systematically the convergence of the onsite and first neighbor self-energy
with cluster size, and compare it to the exact lattice DiagMC result
in the regime where it is available.
In the left panels of Fig.~\ref{fig:SigmaR00} we plot the
(purely imaginary) onsite self-energies of the four inner-core sites
of the even clusters as a function of Matsubara frequency 
for $U/t=4$ and two different temperatures. In the right panels
we plot the (purely real) nearest neighbor self-energy corresponding to
the inter-site correlation between two of the innermost cluster
sites. The overall cluster-size dependence is pronounced in all plots
and stronger for the lower temperature.
In the insets of Fig.~\ref{fig:SigmaR00}, we show 
the linear extrapolation of the data at each Matsubara frequency as a function of
$1/N$. A linear extrapolation is possible
\footnote{
According to spin-wave theory \cite{PhysRevB.40.506}
finite-size corrections 
scale as $1/N$, which may apply to the half-filled Hubbard model
dominated by antiferromagnetic spin fluctuations.
}, 
if we neglect the smallest cluster 
$N\times N=2 \times 2$
.
The extrapolated self-energies are plotted in black.
The error bar combines errors from three sources: the QMC statistics in the quantum impurity solver, 
the convergence of the CDMFT self-consistency loop, and the extrapolation in $1/N$.

At high temperature ($\beta t=5$, upper panels of
Fig.~\ref{fig:SigmaR00}) we compare CDMFT results to DiagMC benchmark data for Matsubara frequencies $i\omega_n$ up to $n=10$ (red color). 
For both the onsite and nearest neighbor self-energies, the extrapolated 
CDMFT results are compatible with the DiagMC data within error bars.
For the onsite self-energy (upper left panel), 
the extrapolated (and DiagMC) self-energy is quite 
different from the largest considered cluster $N\times N=8\times 8$, especially at
the first $n=0$ Matsubara frequency. 
For the nearest-neighbor self-energy, the extrapolation effect extends
to Matsubara frequencies as high as $n=10$. 

At lower temperature ($\beta t=12.5$, lower panels of
Fig.~\ref{fig:SigmaR00}), which is out of reach of the DiagMC algorithm,
the effects of the extrapolation are even stronger, and enhance the impact of the correlations.

\subsection{Center focused extrapolation (CFE) of the self-energy}

After extrapolating the self-energy at the center of the cluster, we approximate the {\it lattice} self-energy in real space by a {\it center focused extrapolation (CFE)} defined in the following.
For each lattice displacement $\mathbf{R}$ with vector elements greater than or equal to zero, we take the large $N$ extrapolation of the cluster self-energy between the central site $\mathbf{R_c}$ (c.f. Fig.~\ref{fig:sketch}) and $\mathbf{R}+\mathbf{R_c}$, i.e.
\begin{equation}
  \Sigma_\text{latt}(\mathbf{R}, i \omega_n) = \Sigma^{N \rightarrow \infty}_{\mathbf{R}+\mathbf{R_c},\mathbf{R_c}}(i \omega_n),
  \label{eq:CFE-R-space}
\end{equation}
For negative displacements we infer the value by using the rotational symmetry of the cluster. This procedure differs substantially from the conventional periodization, where averages over the whole cluster are performed\cite{CDMFTOriginal}.

In practice, given that we solve CDMFT only up to $N\times N=9\times 9$, our strategy is two-fold.
For the short distances $\mathbf{R}$, we use the extrapolated self-energy values computed in the previous subsection.
For the largest $\mathbf{R}$ however, where no $1/N$ extrapolation is possible given the lack of data points, we use instead the value of the largest cluster.
This approximation is justified because the self-energy in real space  
$\Sigma(i\omega_n,\mathbf{R})$ decays on length scales smaller than our largest cluster. 
Indeed in Fig.~\ref{fig:self_energy_vs_R} we show $\Sigma(i\omega_1,\mathbf{R})$
for $\beta t = 5$ and different values of the interaction $U/t$. Even for the
largest interaction value $U/t=4$ do we observe a decay of the absolute value
of the self-energy at the first Matsubara frequency to zero within less than
$5$ lattice constants.

In Fig.~\ref{fig:CDET-CDMFT} we compare the Fourier transform
\begin{equation}
  \Sigma_\text{latt}(\mathbf{k}, i\omega_n) = \sum_{\mathbf{R}} \Sigma^{N \rightarrow \infty}_{\mathbf{R}+\mathbf{R_c},\mathbf{R_c}}(i \omega_n) \, \mathrm{e}^{-\mathrm{i} \mathbf{k} \cdot \mathbf{R}}
\end{equation}
of this estimator of the lattice self-energy (black) with the exact DiagMC computation (green)
for the first two Matsubara frequencies ($\omega_{n=\{0,1\}}$), at the temperature where it is available.
We further compare these to the self-energy obtained by using the same procedure directly
on the $N\times N=8\times 8$ results, i.e. without the large $N$ extrapolation (blue dashed),
and to the ``standard'' periodization of the $N\times N=8\times 8$ self-energy (red) \cite{CDMFTOriginal}.
We find again an excellent agreement, within error bars, between the CDMFT+CFE and the exact DiagMC results.
In particular, for the largest available $U$,
we see a substantial improvement over the conventional self-energy periodization, and a
sizable correction of the self-energy data obtained without extrapolation.
We see in the second to lowest panel on the right side ($\beta t=5$, $U/t=4$, $n=0$) that
the CFE procedure significantly improves even details of the
self-energy along the path $\Gamma$-$X$-$M$-$\Gamma$ (i.e. non-monotonous behavior between $X$ and $M$ points).
\begin{figure}
\centering
\includegraphics[width=0.49\textwidth]{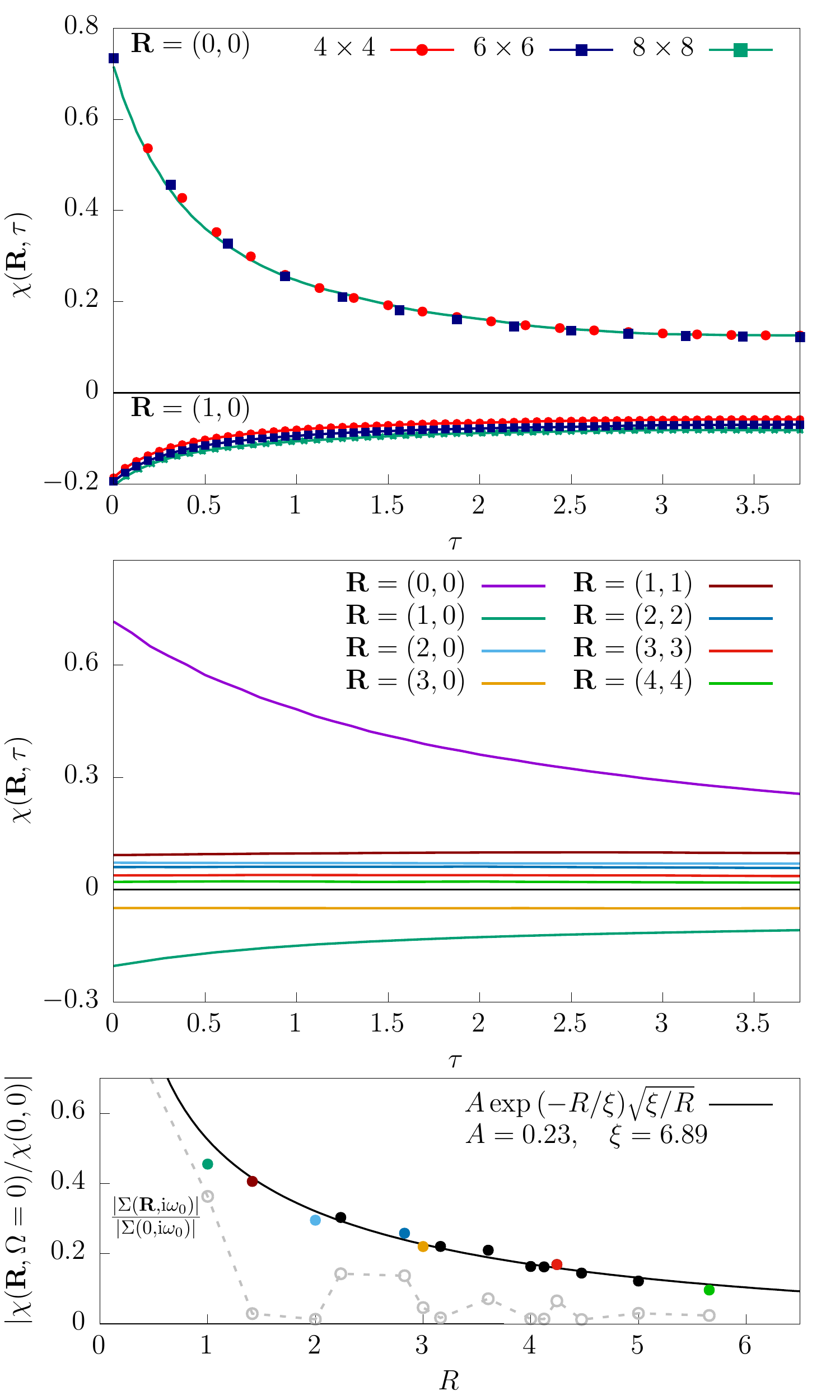}
\caption{Upper panel: Spin-spin correlation function $\chi({\bf R},\tau)$ as
  defined in Eq.~\ref{eq:spin-susceptibility} at distances $R=0$
  and $R=1$, for $\beta t = 7.5, \, U/t = 4$, and three different cluster sizes.
  While the onsite component appears to be almost cluster-size
  independent, the nearest-neighbor one exhibits a dependence on
  $N$. 
  Central panel: The same as the upper panel, but for fixed cluster size $N\times N=8 \times 8$
  and various horizontal and diagonal displacement vectors.
  Lower panel: Normalized static spin susceptibility (colored) and self-energy (gray) as a function
  of the real-space distance $R$ for an $N\times N=8 \times 8$ cluster for the same displacement 
  vectors as above. An exponential fit of the data is shown in black.
  }
  \label{fig:chi_tau}
\end{figure}

\begin{figure*}
  \centering
  \includegraphics[width=0.99\textwidth]{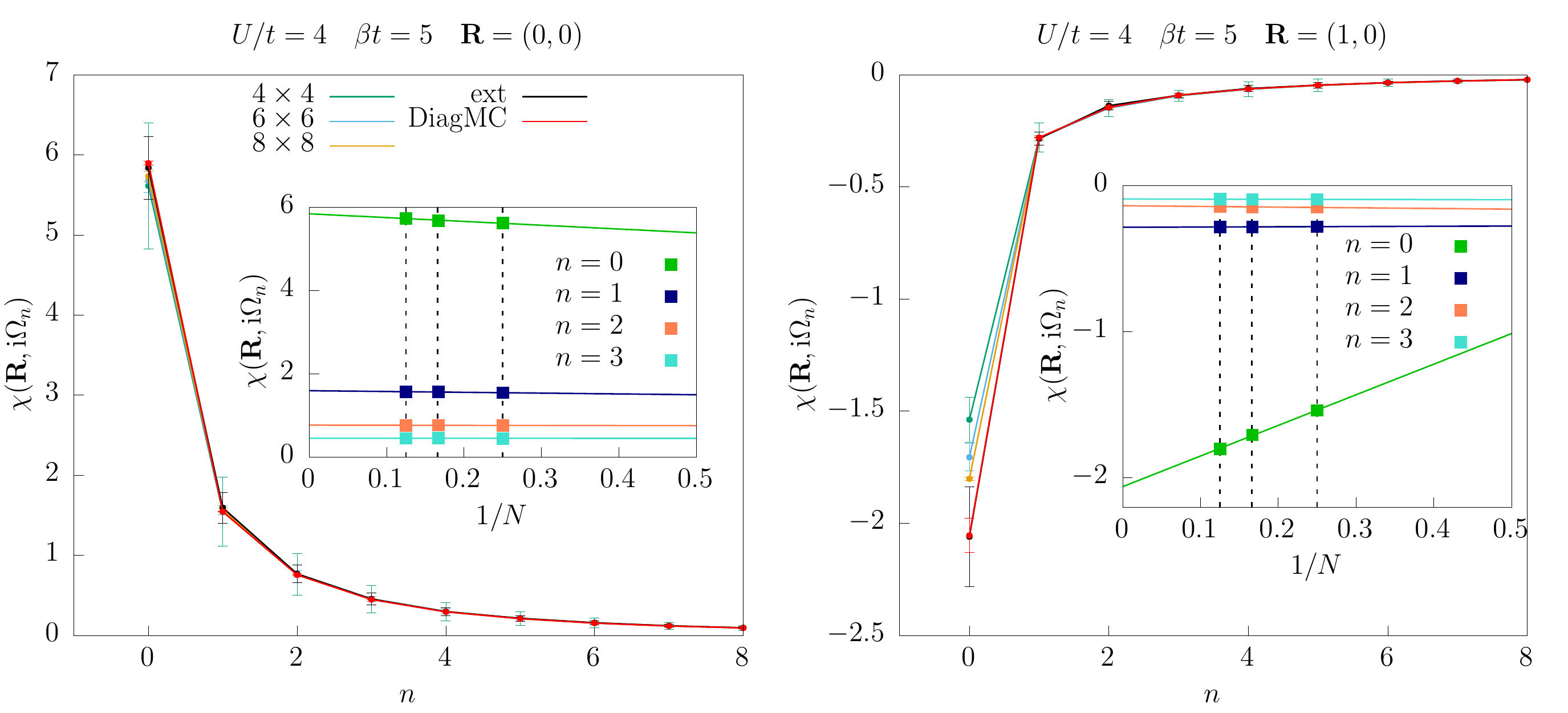}
  \caption{Spin susceptibility for ${\bf R}=(0,0)$ (left) and ${\bf R}=(1,0)$ (right),
    as obtained from the various CDMFT clusters and DiagMC for
    $U/t=4$ and $\beta t=5$. The extrapolation is obtained by fitting
    the data with a $1/N$ linear function, see insets for the fits
    from the first four Matsubara frequencies.
  }
  \label{fig:ChiR00}
\end{figure*}

\subsection{Lattice spectral function of CDMFT+CFE}

The single particle spectra corresponding to the self-energy data of Fig.~\ref{fig:CDET-CDMFT}
are presented in Fig.~\ref{fig:compare-spectra}. In addition we present here also the spectrum
for the cumulant periodization scheme applied to the $N\times N=8 \times 8$ cluster (see Appendix~\ref{app:reper}).

The qualitative differences between the CFE results and the conventional schemes
using data without extrapolation are sizable and most visible around the
Fermi level. The gaps around $X=(0,\pi)$ and between $M$ and $\Gamma$
at $(\pi/2,\pi/2)$ are more pronounced due to the larger value of the
CFE self-energies. 
This shows that the CFE has a qualitative impact on computed observables, 
even at high temperature.

%%%%%%%%%%%%%%%%%%%%%%  
%%%%%%%%% MAGNETISM
%%%%%%%%%%%%%%%%%%%%%%
\section{Spin-spin correlation function}
\label{sec:spinsusc}

Let us now consider the spin susceptibility of the cluster
\begin{align}
  \label{eq:spin-susceptibility}
  \chi({\bf R}, \tau) \equiv &\bigl\langle[\hat{n}_{\uparrow}({\bf R + R_c},\tau) - \hat{n}_{\downarrow}({\bf R + R_c},\tau)]\nonumber\\
  &\times[\hat{n}_{\uparrow}({\bf R_c},0)- \hat{n}_{\downarrow}({\bf R_c},0)]\bigr\rangle\;.
\end{align}
The computation of the full lattice susceptibility, which uses
the impurity two-particle vertex function as an input to the lattice Bethe-Salpeter equation\cite{DMFTreview},
is currently computationally too demanding for large clusters.
However, in the limit of infinite cluster size, we expect the two quantities to coincide.
In Fig.~\ref{fig:chi_tau} (upper panel), we show the imaginary time onsite and nearest-neighbor 
spin-spin correlation functions for $U/t=4$ and $\beta t=7.5$, for different cluster sizes. 
For these parameters both components are i) sizable, ii) fully dynamic (i.e. beyond static
mean-field approaches), and iii) antiferromagnetic (as expected for the present
model). We see a quick convergence of the onsite susceptibility with $N$.
The nearest-neighbor component for $R=1$ has a stronger cluster dependence.
In the central panel of Fig.~\ref{fig:chi_tau} we show
the spatially resolved dynamical susceptibility $\chi ({\bf R}, \tau)$ for the $8 \times 8$ cluster. 
Remarkably, the dynamic part of the susceptibility seems to 
decay much faster with distance than the static part (i.e. already
$\chi(\mathbf{R}=(1,1), \tau)$ is barely dependent on $\tau$). 
This could indicate the presence of two
length scales, one for coherent (i.e. dynamic) and one for non-local
correlations that can be captured by static mean fields. 
The real-space decay of the static
(i.e. $\tau$-integrated) spin susceptibility $\chi({\bf R},\Omega=0)$ 
is presented on the bottom panel of Fig.~\ref{fig:chi_tau}
and compared to that of the corresponding self-energy at the
smallest Matsubara frequency.
The susceptibility data is well approximated by an Ornstein-Zernike
form $A \exp(-R/\xi)\sqrt(\xi/R)$ with a correlation length of $\xi=6.89$.
The fact that we manage to capture the exponential decay of spin-spin
correlations reaching far beyond the scale of the clusters under consideration
is indeed quite remarkable.
A similar analysis for the self-energy proves to be less feasible, given the
fact that it exhibits a much stronger radial dependence.
We observe, however, that the self-energy decays overall on smaller length scales than the susceptibility. 

In order to assess the validity of these susceptibility results we compare it for different
cluster sizes $N\times N = 4\times 4, 6\times 6, 8\times 8$ to the DiagMC benchmark in Fig.~\ref{fig:ChiR00}, where plot $\chi({\bf R}, i\Omega_n)$ as a function of Matsubara frequency for the two
displacements ${\bf R}=(0,0)$ (left) and ${\bf R}=(1,0)$ (right).

As for the self-energy, we perform an extrapolation to infinite cluster size
(c.f. insets of Fig.~\ref{fig:ChiR00}) and obtain excellent agreement, within error bars,
with the DiagMC data. We observe a strong dependence on $N$ only for the displacement
${\bf R}=(1,0)$ at the first Matsubara frequency.

Our analysis of the spin-spin susceptibility suggests that real-space
cluster approaches such as CDMFT are valid even at temperatures and/or interaction values
for which the magnetic correlation length exceeds the cluster size,
as long as the self-energy decay in real-space is sufficiently captured.
However, we expect that the feedback of this long range magnetic mode
onto the electronic self-energy is not correctly described at this cluster size.

%##############################################################
%##############################################################

\section{Conclusion}
\label{sec:conclusion}

In this work, we have revisited the CDMFT calculations for the single band
Hubbard model on the 2D square lattice at half filling.  We performed a
detailed momentum and real-space analysis of the spectral properties for
different cluster sizes of up to $9 \times 9$ sites.  Using a systematic
benchmark with the exact DiagMC result at the lowest temperature for which it
is obtainable, we have shown that an approximation scheme of the lattice self-energy based on
the center of the cluster is superior to the conventional periodization
approaches based on averages over the cluster. In fact, the broken translational
symmetry of the cluster Green's function and self-energy is simply a manifestation
of the bulk-like nature of the central cluster sites, making them the proper
basis for the approximation of the respective lattice quantities.
We have further shown that the exponential decay of spin-spin correlations is very well
captured by CDMFT calculations, even when correlations extend far beyond the size of
the cluster.
Finally, CDMFT calculations and the CFE extrapolation can be carried out to 
temperatures currently inaccessible to exact diagrammatic Monte Carlo techniques,
making the CDMFT+CFE procedure a powerful computational tool to access the physics
of non-local correlations beyond dynamical mean-field theory.

\section{Acknowledgments}
The authors thank A. Georges, K. Held, J. LeBlanc, A. Toschi,
G. Sangiovanni and in particular A.-M. Tremblay for very valuable
discussions, and S. Heinzelmann and C. Hille for a critical reading of
the manuscript. 
The Flatiron Institute is a division of the Simons Foundation.
We acknowledge financial support from the Deutsche
Forschungsgemeinschaft (DFG) through ZUK 63 and Project No. AN
815/6-1. The present work was also supported by the Austrian Science
Fund (FWF) through the Erwin-Schr\"odinger Fellowship J 4266 - "{\sl
  Superconductivity in the vicinity of Mott insulators}'' (SuMo,
T.S.), as well as the European Research Council for the European Union
Seventh Framework Program (FP7/2007-2013) with ERC Grant No. 319286
(QMAC, T.S.) and the Simons Collaboration on the Many-Electron
Problem.  Calculations were partly performed on the cluster at the MPI
for Solid State Research in Stuttgart.  The authors acknowledge also
the CPHT computer support team for their help.

\appendix

\section{Antiferromagnetic order}
\label{app:Neel}
In this appendix, we consider antiferromagnetic (AFM) ordering in our CDMFT computations.

\begin{figure}[ht]
  \centering
  \includegraphics[width=0.49\textwidth]{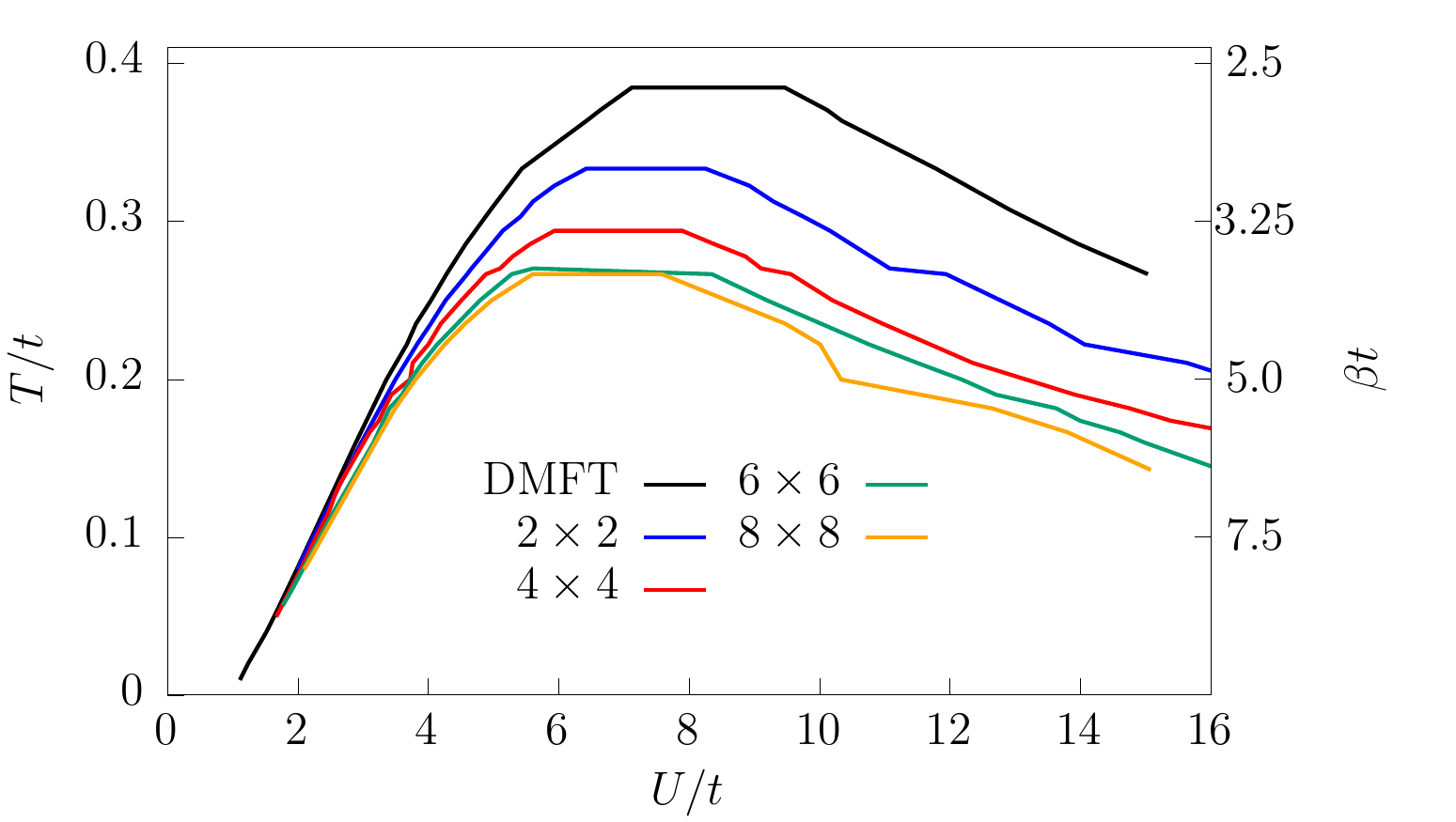}
  \caption{N\'eel temperatures for a DMFT and various CDMFT
    calculations with even cluster sizes $N$. }
  \label{fig:Neel_even}
\end{figure}

In Fig.~\ref{fig:Neel_even} we provide the phase boundary $T_N(U)$ as
determined by the CDMFT with the criterion:
AFM order for $m_\text{AFM} \geq 0.1$ and PM one for
$m_\text{AFM} \leq 0.1$. We consider cluster sizes up to $8 \times 8$. 
Due to commensurability with the AFM checkerboard symmetry, we restrict
our analysis to even $N$ (even though odd clusters could also be
easily accessed in self-consistent cycles analogous to AFM single site
DMFT).  We observe a monotonous reduction of the N\'eel temperature
with increasing cluster size, with a pronounced
$U$-dependence: the reduction of $T_N$ in the strong coupling regime
is significantly stronger compared to the small $U$ region.
This reduction is expected, and compatible with the exact N\'eel temperature of the 2D square lattice, which is known to be
$T_N=0$ from the Mermin-Wagner theorem~\cite{mermin-wagner}.

\label{section:magnetism}

\section{Re-periodized spectral functions}
\label{app:reper}
\begin{figure}[t]
  \centering
  \includegraphics[width=0.45\textwidth]{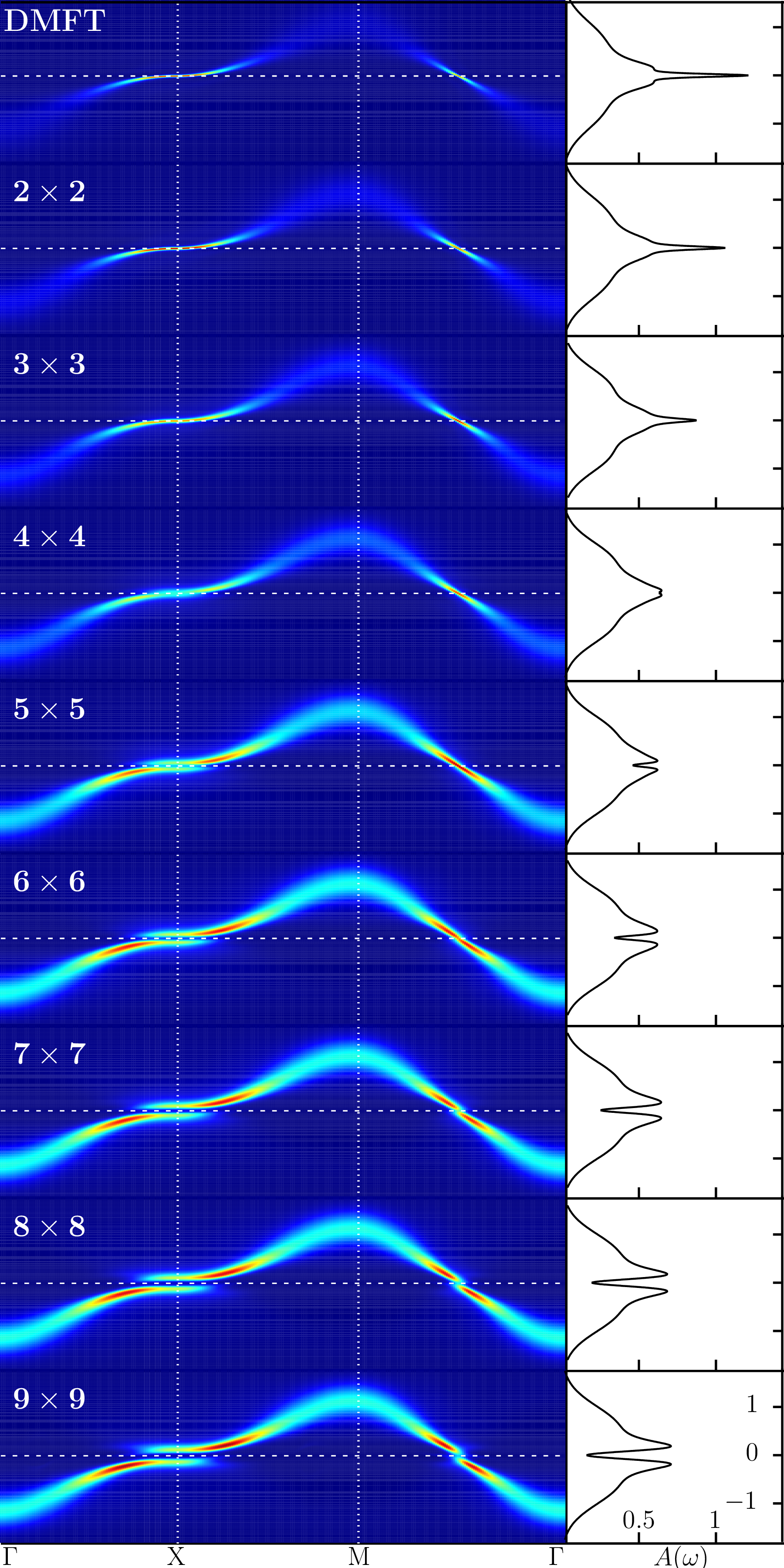}
  \caption{Momentum resolved spectral functions along a closed path
    inside the first Brillouin zone (left) and full spectral function
    of the local Green's function (right), for $U/t=4$, inverse
    temperature $\beta t = 12.5$, and different square clusters, as
    obtained by using the cumulant periodization. The dotted line
    corresponds to $\omega =0$.  While the single-site DMFT ($N=1$) is
    clearly in the metallic phase, the quasiparticle peak vanishes
    with increasing $N$, and $N\times N=9 \times 9$ is already close to the
    insulating phase characterized by a gap in the local spectral
    function.}
  \label{fig:spectrak}
\end{figure}

\begin{figure}[t]
  \centering
  \includegraphics[width=0.49\textwidth]{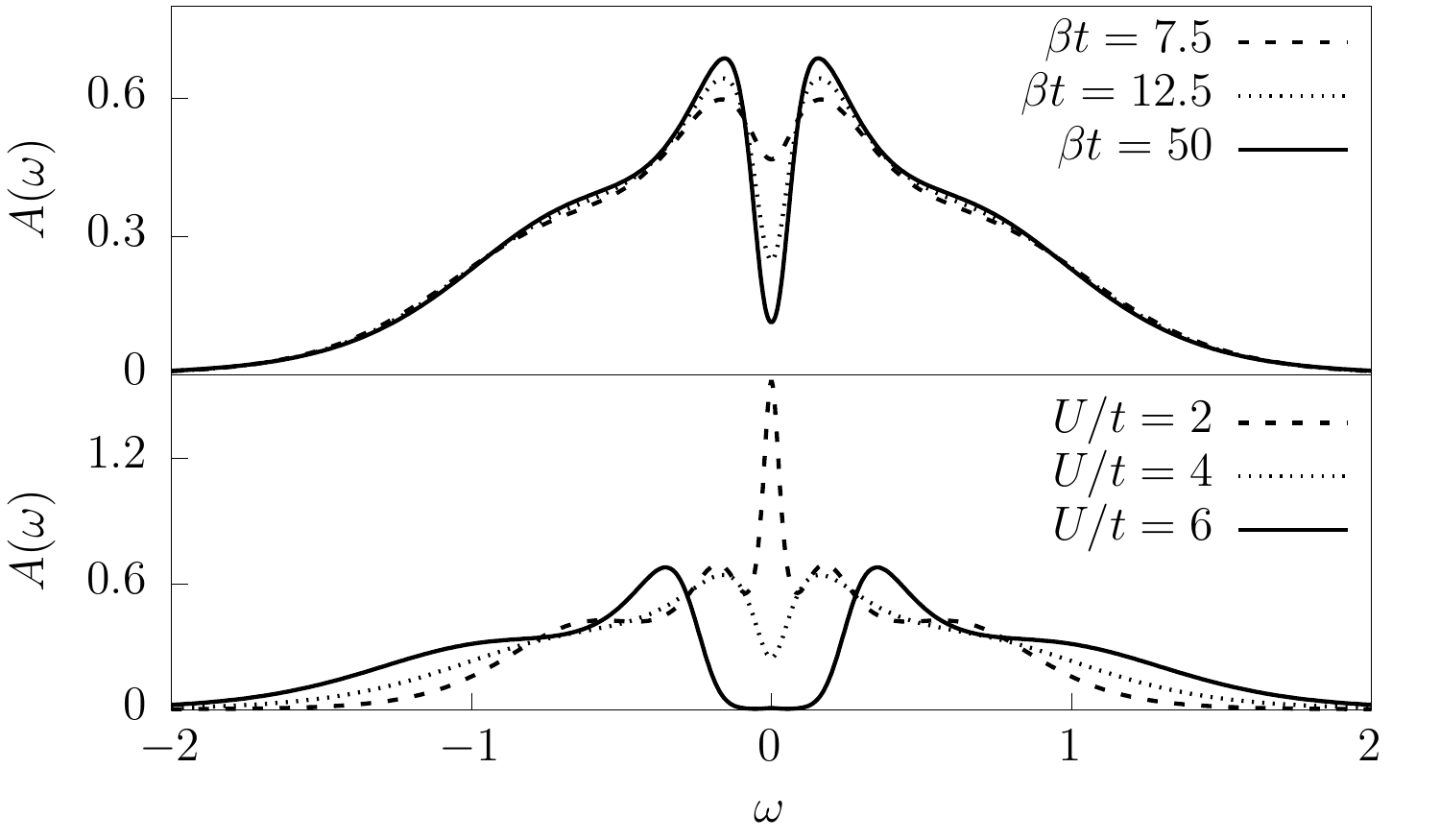}
  \caption{Spectral function of the local lattice Green's function for
    $N\times N=7 \times 7$, for different inverse temperatures and fixed
    $U/t=4$ (upper panel), and for different values of $U$ and a fixed
    inverse temperature $\beta t =12.5$ (lower panel).}
    \label{fig:mit2}
\end{figure}

\begin{figure}[h]
  \centering
  \includegraphics[width=0.49\textwidth]{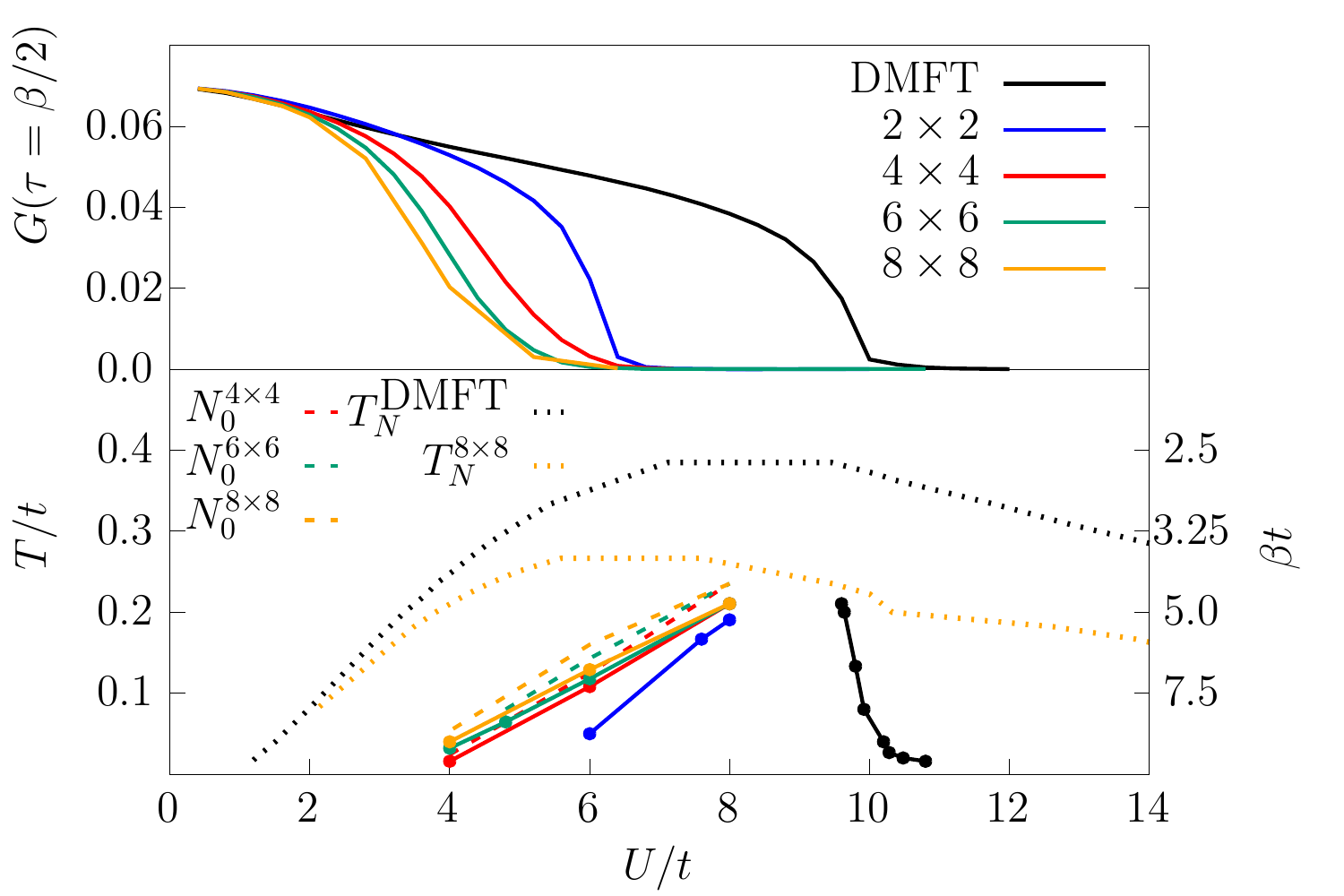}
  \caption{Weight of the imaginary time local lattice Green's function
    evaluated at $\tau = \beta/2$, for a fixed inverse temperature
    $\beta t = 12.5$ and different cluster sizes (upper panel) and
    phase diagram with the Mott transition line for different cluster
    sizes (lower panel). For comparison, we show also the transition
    lines as obtained from the local lattice Green's function (dashed
    lines). The N\'eel temperatures for the smallest and largest
    cluster size is indicated by dotted
    lines.}
    \label{fig:mit3}
\end{figure}

In this supplemental section we present directly postprocessed CDMFT
data without extrapolation for various cluster sizes.  In order to
compute momentum dependent spectral functions in the Brillouin zone of
the 2D square lattice we follow the cumulant periodization scheme of
Ref.~\onlinecite{sakai-CDMFT}.  Instead of periodizing the self-energy
directly, its cumulant
$M (\ii \omega_n)= (\ii \omega_n + \mu - \Sigma(\ii \omega_n) )^{-1}$
is periodized to obtain the lattice Green's function
$G(\mathbf{k},\ii \omega_n)$ by
\begin{equation}
  \label{eq:lattice-quantity}
  M(\mathbf{k}, \ii \omega_n) = \frac{1}{N} \sum_{i,j=1}^{N} M_{i,j}(\ii \omega_n) \, \ee^{\ii \mathbf{k} (\mathbf{R}_i - \mathbf{R}_j)} \;,
\end{equation}
with $\mathbf{R}_i$, $\mathbf{R}_j$ being the real-space positions of
the sites $i$ and $j$ in the cluster, and the origin in
$\mathbf{R}=(0,0)$ corresponding to the central site for $N$ odd and
to one of the innermost sites for $N$ even.
We note that also the application of this scheme turned out to be
problematic in some situations - which actually motivates the present
analysis to account for the real-space anisotropies of CDMFT - but is
used here as a reference since it was shown to perform better than
re-periodizations of the self-energy or the Green's
function~\cite{sakai-CDMFT}.

In Fig.~\ref{fig:spectrak} we show results for the single-particle
spectral function as obtained from CDMFT for $U/t=4$ and an inverse
temperature $\beta t=12.5$ by employing~\eqref{eq:lattice-quantity}
and subsequent analytical continuation with the Maximum Entropy
method~\cite{maxent-bryan,Manuel_maxent}.
Each of the nine horizontal panels contains a momentum resolved
intensity map for $A(\omega,\mathbf{k})$ (left) on the path
$\Gamma$-$X$-$M$-$\Gamma$ in the cubic Brillouin zone and the
$\mathbf{k}$-integrated local spectral function $A(\omega)$ (right).
The evolution from top (``DMFT'') to bottom (``$9 \times 9$'') illustrates the
cluster size dependence of the spectrum at fixed interaction and temperature.
As function a of $N$, we observe an MIT transition which reflects in
$A(\omega)$ as a spectral weight transfer from sharp quasiparticle
excitations around the Fermi-level $\varepsilon_F$ to incoherent
Hubbard bands, in agreement with previous studies on smaller cluster
sizes~\cite{park-CDMFT,olivier-CDMFT}. However, the opening of the gap
in $A(\omega)$ is a gradual process crossing a pseudogap-like regime
between $N\times N=4 \times 4$ and $N\times N=7 \times 7$. This effect
originates from a momentum selective opening of the gap (see also
Ref.~\onlinecite{PhysRevB.86.155109}). Closer inspection of
$A(\omega,\mathbf{k})$ for the intermediate regime of
$N\times N = 5 \times 5$ and $6 \times 6$ clearly reveals the absence
of a Fermi surface at the antinodal point $X=(\pi,0)$, while the
quasiparticle band crossing $\varepsilon_F$ in the nodal direction
around $(\pi/2,\pi/2)$ remains sharp.

To complement our analysis of the MIT we also study the $U$ and
$\beta$ dependencies of $A(\omega)$. Figure~\ref{fig:mit2} shows
$A(\omega)$ for $N\times N=7 \times 7$ and $U/t=4$. The temperature
dependence, especially of the spectral weight at $\varepsilon_F$,
indicates the presence of a fully developed gap (in agreement with the
corresponding $N\times N=7\times 7$ momentum resolved spectrum) which is
barely closed by incoherent thermal smearing. Data without analytical
continuation are reported in the lower panel of Fig.~\ref{fig:mit3},
with $G(\tau = \beta/2) = A(\omega = 0)$. 

Here our results for the MIT as a function of $N$, $U$, and
temperature are summarized. Remarkably, the slope of the $T_c$
vs. $U_c$ transition line is inverted with respect to the single-site DMFT already for the
smallest $N\times N=2 \times 2$ cluster, in agreement with previous
works~\cite{olivier-CDMFT}. Increasing $N$ leads to a simultaneous
increase of $T_c$ at fixed $U$ and decrease of $U_c$ at fixed $T$
which is consistent with previous calculations directly in the
thermodynamic limit,~e.g. D$\Gamma$A~\cite{DGA4,DGA5} and
DiagMC~\cite{CDET,old_calc}, where the critical value shows the
behavior $U_c \to 0$.

\bibliography{references.bib}

\end{document}